# Anomalous switching pattern in the ferrimagnetic memory cell

Zhuo Xu[1], Zhengping Yuan[1], Xue Zhang[1], Zhengde Xu[1], Yixiao Qiao[1], Yumeng Yang[1,2], Zhifeng Zhu[1,2]†

[1]School of Information Science and Technology, ShanghaiTech University, Shanghai, China 201210

[2]Shanghai Engineering Research Center of Energy Efficient and Custom AI IC, Shanghai, China 201210

**Abstract**

Replacing the ferromagnet with ferrimagnet (FiM) in the magnetic tunnel junction (MTJ) allows faster magnetization switching in picoseconds. The operation of a memory cell that consists of the MTJ and a transistor requires reversable magnetization switching. When a constant voltage is applied, we find that the spin-transfer torque can only switch the FiM-MTJ from parallel to antiparallel state. This stems from the small switching window of FiM and the dynamic resistance variation during the magnetization switching. We find the resulting current variation can be suppressed by reducing the magnetoresistance ratio. Furthermore, we demonstrate that the switching window can be expanded by adjusting the amount of Gd in FiM. We predict that the polarity of both switching current ($J_{c,\text{switch}}$) and oscillation current ($J_{c,\text{osc}}$) reverses at the angular momentum compensation point but not the magnetization compensation point. This anomalous dynamic behavior is attributed to the different physical nature of magnetization switching and oscillation in FiM, which must be considered when designing FiM-based MRAM.

**Introduction**

In recent years, rapid advancements in big data and artificial intelligence have created a high demand for improved memory technology that has lower latency, larger storage capacity, and reduced power consumption. The conventional memory technologies, such as SRAM, DRAM, and Flash, are very difficult to meet these requirements [1]. In contrast, the magnetic random-access memory (MRAM) that can be manipulated by spin-transfer torques (STT) [2-4] or spin-orbit torque (SOT) [5-15] is found to be very promising due to its inherent nonvolatility, fast switching speed below 5 ns [16], low power consumption in 0.15 pJ [17-21], small footprint below 5 nm [22], and large capacity over 1 GB [23]. A typical memory cell of MRAM consists of a magnetic tunnel junction (MTJ) and a transistor [24]. By applying an appropriate bias voltage to the memory cell, current flows through the MTJ and switches the magnetization of the ferromagnet (FM) layer through the STT. The switching speed of the MRAM is limited by the internal field of the ferromagnetic free layer [25]. By replacing the FM with antiferromagnet (AFM) [26, 27] or ferrimagnet (FiM) [28], the strong exchange field comes into play, resulting in a faster switching in picoseconds [29-33]. Compared to AFM, FiM [34-39] has unique properties such as the nondegenerate magnetic and angular momentum compensation point [40-47]. These additional freedoms allow further performance improvement in designing novel functional devices [48-56].

The successful writing of the memory cell includes writing both 1 and 0, which is represented by the antiparallel (AP) and parallel (P) of magnetization in the free layer (FL) and the pinned layer (PL). In the FM-based MTJ, it has been found that the use of a constant voltage source produces different magnetization dynamics compared to that with a constant current source [57]. In the constant voltage mode, the dynamic change of magnetization during the switching leads to a varied resistance of MTJ. Due to the voltage division between MTJ and transistor, the current flowing through the memory cell is also changed dynamically, which deteriorates the writing performance. In principle, the same problem will also be manifested in the FiM-based memory. Given the distinctive properties of FiM, such as the impact of their two compensation points on switching, it is both theoretically significant and practically essential to comprehensively evaluate the writing performance of FiM-based memory. For instance, compared to the previous work [58], our ferrimagnetic MTJ permits ultrafast switching in 0.108 ns when a current of 750 μA is applied. In addition, our analysis reveals that the angular momentum compensation plays a crucial role in determining the switching and oscillation dynamics, which show opposite trend as their underlying physics in FiM is different from that in FM. This contrasting behavior introduces a level of complexity that must be carefully considered when optimizing the performance of FiM-based memory cells.

**Method**

In this work, we study the FiM alloy, $Gd_xFeCo_{(1-x)}$, which is commonly used in experiments[45, 59-62]. The structure of FiM-MTJ is illustrated in Fig. 1(a), where the FiM and FM are used as the FL and PL, respectively. Depending on the direction of $\mathbf{J}_c$ that flows through the MTJ, the damping-like torque (DLT) of STT changes the magnetization of the FiM layer between P and AP states. Following Ref. [41], the definition of P and AP state in FiM MTJ refers to the net magnetic moment, whereas the high and low resistance is determined by the relative orientation between $\mathbf{m}_{FeCo}$ and the magnet moment of pinned layer ($\mathbf{m}_{PL}$), i.e., high resistance corresponds to opposite alignment of $\mathbf{m}_{FeCo}$ and $\mathbf{m}_{PL}$. The magnetization dynamics of the two sublattices in the FiM are obtained by solving the coupled Landau–Lifshitz–Gilbert–Slonczewski (LLGS) equations [63-65]:

$$\frac{d\mathbf{m}_i}{dt} = -\gamma_i \mu_0 \mathbf{m}_i \times \mathbf{H}_{\text{eff},i} + \alpha_i \mathbf{m}_i \times \frac{d\mathbf{m}_i}{dt} - \gamma_i \mu_0 \mathbf{m}_i \times (\mathbf{m}_i \times \mathbf{H}_i^{\text{DLT}}). \quad (1)$$

In this equation, $i$ denotes the sublattice FeCo or Gd. $\gamma_i = g_i \mu_B / \hbar$ is the gyromagnetic ratios with $g_{FeCo} = 2.2$ and $g_{Gd} = 2$ [28]. $\alpha_{FeCo} = \alpha_{Gd} = 0.015$ [28] is the damping constant. $\mathbf{H}_{\text{eff},i} = \mathbf{H}_{k,i} + \mathbf{H}_{\text{ex},i}$ is the effective magnetic field, in which the anisotropy field $\mathbf{H}_{k,i} = (2K_{u,i}/M_{s,i})m_{z,i}\mathbf{z}$ with $K_{u,FeCo} = K_{u,Gd} = 1\times10^5$ J/m³ [28] and the exchange fields $\mathbf{H}_{\text{ex},FeCo} = (J_{\text{ex}}/((1-x)\mu_{FeCo}))\mathbf{m}_{Gd}$, $\mathbf{H}_{\text{ex},Gd} = (J_{\text{ex}}/(x\mu_{Gd}))\mathbf{m}_{FeCo}$ with the exchange constant $J_{\text{ex}} = 1.09\times10^{-21}$ J [66], magnetic moment $\mu_{FeCo} = 1.92\mu_B$ and $\mu_{Gd} = 7.63\mu_B$ [42, 67]. The damping-like effective field of STT $\mathbf{H}_i^{\text{DLT}} = \eta\hbar J_c \mathbf{m}_p/(2\mu_0 t_{FL} M_{s,i})$, where $\mathbf{m}_p$ is always along the +z direction, the free layer thickness $t_{FL} = 1.2$ nm and the STT efficiency $\eta = P/(1+P^2\cos(\theta_i))$ with $P = 0.4$ [24]. $\theta_i$ denotes the angle between sublattice magnetization and $\mathbf{m}_p$. The saturation magnetization $M_{s,FeCo} = (1-x)\mu_{FeCo}/a^3$ and $M_{s,Gd} = x\mu_{Gd}/a^3$ with the lattice constant $a = 0.352$ nm [67]. Based on these parameters, one can get the angular momentum compensation $x_{AMC} = 0.186$ and the magnetization compensation $x_{MC} =$

0.201. We have developed the numerical model using the hardware programming language Verilog-A to integrate the LLGS equation. This enables us to co-simulate the FiM-MTJ and a 180 nm MOSFET in the circuit design software Cadence Virtuoso.

**Results and discussion**

We first study the magnetization dynamics in the MTJ stack driven by a current source, which offers a clear physical picture without including the complex effect from MOSFET. As shown in Fig. 1(b), the initial state at $J_c = 0$ has a net magnetization pointing in −z direction. When a positive $J_c$ is applied, $\mathbf{m}_{FeCo}$ is switched to +z direction (region 2) since $\mathbf{m}_p = +z$, and $\mathbf{m}_{Gd}$ is aligned to −z direction due to the strong exchange interaction between the sublattices. When $J_c$ is further increased, the sublattices develop into oscillation (region 3 and 4), and finally both align with $\mathbf{m}_p$ (region 5) [64, 68]. This phase diagram is unique to FiM, in which the switching happens before oscillation. In contrast, oscillation happens before switching in FM [63]. We will demonstrate that this difference is crucial for understanding the switching and oscillation behaviors in FiM. For the negative $J_c$, there is no switching region [64, 68], and the rest of phase diagram is the same as that with the positive $J_c$. Similarly, the phase diagram for the MTJ initially in the P state is shown in Fig. 1(c). It is worth noting that the switching region, defined as the current range between $J_{c,switch}$ and $J_{c,osc}$, is very small. As we will discuss later, this produces challenges in achieving successful memory writing when MOSFET is introduced.

As illustrated in the inset of Fig. 2(a), the memory cell studied here consists of the

FiM-MTJ and an NMOS transistor. The transistor will be conducting when $V_{GS}$ exceeds the threshold voltage ($V_{th}$) and $V_{DS}$ exceeds $V_{GS}-V_{th}$. In our study, a relatively large $V_G$ is chosen to provide a high current that can be used to switch the magnetization. Initially, the MTJ is in the P state. When the bit-line ($V_{BL}$) is connected to a positive voltage and the source-line ($V_{SL}$) is grounded, current flows from up to down, which would switch the MTJ to AP state for a sufficient large $J_c$. In Fig. 2(a), no switching happens since the small $J_c$ does not provide sufficient STT to overcome the Gilbert damping. When $V_{BL}$ is increased to 1.25 V, $\mathbf{m}_{FeCo}$ is successfully switched from +z to −z as shown in Fig. 2(b). For a reversable switching from AP to P, the bias voltage also needs to be reversed, i.e., $V_{SL}$ is connected to a positive voltage source and $V_{BL}$ is grounded. Note that in this case, the source terminal of MOSFET changes to point A [see inset of Fig. 2(c)]. Due to the voltage division between the MTJ and MOSFET, $V_{GS}$ becomes smaller than $V_G$. This is known as the source degeneration [1], and it makes the switching voltage higher than the other case. This can be observed in Fig. 2(c) where $V_{SL}$ = 1.98 V, which is already larger than the switching voltage in Fig. 2(b), is not sufficient to excite magnetization reorientation. When $V_{SL}$ is gradually increased, we found the non-switching region [Fig. 2(c)] changes to the oscillation region [Fig. 2(d)], and there is no switching region in between.

To understand the oscillation, we need to look at the current change during the transition process. Different from the system under a constant current [cf. Fig. 1(a)], in the memory cell driven by a constant voltage [cf. Fig. 2], the current flowing through MTJ changes dynamically since the MTJ resistance ($R_{MTJ}$) is a function of $\mathbf{m}$, i.e., $R_{MTJ}$

= $R_P+(R_{AP}-R_P)(1-\cos\theta)/2$, where $\theta$ is the angle between $\mathbf{m}_{FeCo}$ and $\mathbf{m}_p$. During the AP to P switching, $R_{MTJ}$ becomes smaller, resulting in an increase in the current. As shown in the right **y** axis of Fig. 2(d), $J_c$ increases from $4.12\times10^{11}$ to $6.36\times10^{11}$ A/m². Referring to Fig. 1(b), $J_c = 4.12\times10^{11}$ and $6.36\times10^{11}$ A/m² locates in the non-switching and oscillation region, respectively. This explains the oscillation observed in Fig. 2(d). For the parameters used in this study, we find the increase of $J_c$ during the AP to P transition will always enter the oscillation, preventing the occurrence of successful switching. As a comparison, $J_c$ is reduced during the switching from P to AP state [see Fig. 2(b)], which guarantees a successful switching as long as the starting current in the P state is located in the switching region. We want to emphasize that the difference in the reversable switching observed here cannot be explained without considering the MOSFET.

Since the oscillation is induced by the increase in current as the resistance is reduced from AP to P state, one can adjust the resistance change to locate $J_c$ of the P state into the switching region. This can be achieved by reducing the tunneling magnetoresistance (TMR) ratio whereas maintaining $R_P$, which results in a smaller $R_{AP}$, and thus a smaller $J_c$ in the beginning. Figs. 3(a) and 3(b) show the results after reducing the TMR ratio to 50%. Although the final $J_c$ becomes smaller, i.e., $5.97\times10^{11}$ A/m² as shown in Fig. 3(b), successful switching is not achieved. As shown in Fig. 3(c), further reducing TMR ratio to 35% enables the successful switching to P state, where the final $J_c = 5.5\times10^{11}$ A/m² is in the switching region. Fig. 3(d) shows the critical current density that induces switching or oscillation as a function of TMR ratio. The critical current density at the boundary that separates the switching and oscillation region is $5.8 \times 10^{11}$ A/m$^2$, which is identical to that in Fig. 1(b). This clearly demonstrates that the unsuccessful switching is induced by the rise of $J_c$ into the oscillation region. In addition, the nonlinear trend also showcases the important role of the transistor, which is discussed in APPENDIX A.

Although the reversible switching can be achieved by reducing the TMR ratio, it deteriorates the memory performance by making the reading more challenging. From another perspective, the undesirable oscillation can be avoided if the narrow switching region shown in Fig. 1(b) can be expanded. This can be achieved by tuning $x$ which utilizes the unique property of FiM alloy. The change in $x$ leads to a different $M_{s,\text{FeCo}}$ and $M_{s,\text{Gd}}$ that results in a different magnetization dynamics. Fig. 4(a) shows the switching region as a function of $x$. Starting from $x = 0.15$, we find that the switching window reduces linearly when $x$ is increased. This trend remains until $x=x_{\text{AMC}}$, where no successful switching is observed. After that, the switching region becomes larger when $x$ is further increased. We note that $x_{\text{MC}}$ has no effect in this case, whereas it is $x_{\text{AMC}}$ acts as the critical point that separates the two regions with opposite trend. To understand why the transition occurs at $x_{\text{AMC}}$, we first study the switching polarity in three regions, i.e., $x<x_{\text{AMC}}$, $x_{\text{AMC}}<x<x_{\text{MC}}$, and $x>x_{\text{MC}}$. The initial state for all cases is set as the AP state, i.e., $\mathbf{m}_{\text{net}}$ is antiparallel to $\mathbf{m}_p$. For $x<x_{\text{AMC}}$ and $x_{\text{AMC}}<x<x_{\text{MC}}$, $\mathbf{m}_{\text{FeCo}}$ is the dominant magnetization. As shown in Figs. 4(b) and 4(c), under appropriate bias voltage, $m_{z,\text{FeCo}}$ is successfully switched from −1 to 1. In contrast, $\mathbf{m}_{\text{Gd}}$ is the dominant magnetization in the sample with $x > x_{\text{MC}}$, and Fig. 4(d) shows $m_{z,\text{Gd}}$ switches from −1 to 1 with $V_{SL}>0$ and $V_{BL}=0$. Although successful magnetization switching is observed in all three cases, it is worth noting that the switching polarity is opposite between $x_{AMC} < x < x_{MC}$ and the other two cases, i.e., the current need to flow in the opposite direction to switch $\mathbf{m}_{net}$ from AP to P. This is reminiscent of the transition at $x_{AMC}$ as shown in Fig. 4(a). The switching polarities in the three cases can be understood by analyzing the switching of both magnetization and angular momentum. As shown in the inset of Fig. 4(b), current flow from $V_{SL}$ to $V_{BL}$ produces $\boldsymbol{\sigma}$ in +z direction, which switches $\mathbf{m}_{net}$ (solid arrow) to +z direction. Due to the antiparallel orientation between $\mathbf{m}$ and $\mathbf{s}$, we can get that $\mathbf{s}_{net}$ (dot arrow) is switched to −z direction. However, it should be noted that the nature of STT is the transfer of angular momentum [69]. Therefore, in the context of Fig. 4(b), it is the $\mathbf{s}_{net}$ that is switched by STT to −z direction, and the direction of $\mathbf{m}_{net}$ can be determined by analyzing $\mathbf{m}$ and $\mathbf{s}$ of each sublattice. Similarly, for the sample with $x_{AMC} < x < x_{MC}$, the analysis shown in the inset of Fig. 4(c) indicates that only the current flowing from $V_{BL}$ to $V_{SL}$ can switch the magnetization.

Although the reversed polarity at $x_{AMC}$ is explained by the transfer of spin angular momentum, we find the linear trend shown in Fig. 4(a) presents more complexity. Based on the phase diagram shown in Fig. 4(a), we first extract $J_{c,switch}$ and $J_{c,osc}$. As shown in Fig. 4(e) and 4(f), when $x$ is increased towards $x_{AMC}$, $J_{c,switch}$ increases, whereas $J_{c,osc}$ reduces. For $x_{AMC} < x < x_{MC}$, $J_{c,switch}$ becomes negative, which corresponds to the reversed polarity discussed in the last paragraph. It is worth noting that in this region, when $x$ becomes larger, the magnitude of $J_{c,switch}$ reduces, whereas $J_{c,osc}$ increases. When $x$ becomes larger than $x_{MC}$, both $J_{c,switch}$ and $J_{c,osc}$ becomes positive and their trend remains the same compared to the region with $x_{AMC} < x < x_{MC}$. Note that $J_{c,osc}$ firstly decreases, and after across $x_{AMC}$, its magnitude increases as a function of $x$. Conversely, $J_{c,switch}$ demonstrates an inverse trend. This trend is different from the ferromagnet system, which would be independent of $M_s$ when the demagnetizing field is not included. To further investigate the underlying reason for this behavior, we combine the LLGS equation of $\mathbf{m}_{FeCo}$ and $\mathbf{m}_{Gd}$ into an effective magnetic moment [70, 71]:

$$\frac{d\mathbf{M}_{eff}}{dt} = -|\gamma_{eff}|(\mathbf{M}_{eff} \times \mathbf{H}_{eff}) + \frac{\alpha_{eff}\left(\mathbf{M}_{eff} \times \frac{d\mathbf{M}_{eff}}{dt}\right)}{M_{eff}} - \frac{|\gamma_{eff}|(\mathbf{M}_{eff} \times (\mathbf{M}_{eff} \times \mathbf{H}_{STT}))}{M_{eff}}.$$

The analytical expression of $J_{c,switch}$ can be obtained by expanding it into the LL form:

$$\frac{d\mathbf{M}_{eff}}{dt} = -\frac{|\gamma_{eff}|}{1+\alpha_{eff}^2}(\mathbf{M}_{eff} \times \mathbf{H}_{eff}) - \alpha_{eff}\frac{|\gamma_{eff}|}{1+\alpha_{eff}^2}(\mathbf{M}_{eff} \times (\mathbf{M}_{eff} \times \mathbf{H}_{eff})) - \frac{|\gamma_{eff}|}{1+\alpha_{eff}^2}(\mathbf{M}_{eff} \times (\mathbf{M}_{eff} \times \mathbf{H}_{STT})) + \alpha_{eff}\frac{|\gamma_{eff}|}{1+\alpha_{eff}^2}(\mathbf{M}_{eff} \times \mathbf{H}_{STT})).$$

By balancing the Gilbert damping (2nd term on the RHS) with the STT (3rd term on the RHS), we can get $J_{c,switch} = \frac{2\alpha_{eff} e \mu_0 t_{FL}}{\eta \hbar} H_{eff}$. However, note that $\alpha_{eff} = \frac{\frac{\lambda_{RE}}{|\gamma_{RE}|^2} + \frac{\lambda_{TM}}{|\gamma_{TM}|^2}}{\frac{M_{RE}}{|\gamma_{RE}|} - \frac{M_{TM}}{|\gamma_{TM}|}} = \frac{\frac{\lambda_{RE}}{|\gamma_{RE}|^2} + \frac{\lambda_{TM}}{|\gamma_{TM}|^2}}{|s_{net}|}$, $J_{c,switch}$ becomes larger when the sample approaches the angular momentum compensation point, which explains the trend of $J_{c,switch}$ observed in Fig. 4(f).

To explain the opposite trend of $J_{c,osc}$, we refer to the phase diagram shown in Fig. 1(b), where the MTJ is in the AP state under zero $J_c$. A positive $\mathbf{J}_c$ switches it to the P state (region 2) due to $\boldsymbol{\sigma}=+\mathbf{z}$. However, further increasing $J_c$ results in the oscillation (region 3). Note that since $\boldsymbol{\sigma}$ is still along $+\mathbf{z}$, the oscillation is induced by the balance of STT and Gilbert damping on the Gd atom, whereas the FeCo atom impedes the oscillation. As the sample moves away from the angular momentum compensation point [cf. Fig. 4(f)], the FeCo atom becomes more dominant. Therefore, one needs to apply a higher $J_c$ to achieve oscillation. This explains the trend of $J_{c,osc}$ in Fig. 4(f). Overall, the contrasting trends of $J_{c,\,switch}$ and $J_{c,osc}$ can be attributed to the different physical mechanisms, i.e., the switching is determined by the FeCo atom, whereas the oscillation is determined by the Gd atom in the sample where FeCo is the dominant atom.

## Conclusion

We studied the writing performance of the FiM-MTJ ($Gd_{17.8}FeCo_{82.2}$) connected to a 180 nm transistor under a constant voltage. Successful switching from 0 to 1 can always be obtained, whereas the reversed switching cannot be achieved. We attribute this to the small switching window of FiM together with the dynamic voltage division between the MTJ and the transistor. Reducing the TMR ratio partially solves this problem, but the reading performance is deteriorated. We proposed an improved solution that enlarges the switching window by adjusting the Gd composition. We found that both the oscillation current and switching current reverses at $x_{AMC}$. We explain this as the nature of STT is the transfer of angular momentum. In addition, the oscillation current reduces when approaching $x_{AMC}$, whereas the switching current shows an opposite trend. This behavior is unique in FiM since the physics dominating the switching and oscillation depends on different atoms.

†Corresponding Author: zhuzhf@shanghaitech.edu.cn

**Acknowledgements:** This work was supported by National Key R&D Program of China (Grant No. 2022YFB4401700) and National Natural Science Foundation of China (Grants No. 12104301).

APPENDIX A: THE NONLINEAR TREND OF MEMORY CELL.

The MOSFET operates as a switch that modulates the current flow through the MTJ. This switching function is critical in the functioning of the memory cell, as it determines the effective current that reaches the MTJ, thereby influencing its magnetic state transitions. The nonlinear trend observed in our results can be attributed to the saturation behavior of the MOSFET. As the MOSFET approaches saturation, it no longer behaves as a simple linear resistor, but exhibits characteristics that lead to a reduction in the rate of increase in current. From Fig. 5, it can be observed that the curves for the structure with only MTJ are linear, whereas it shows saturation when the MOSFET is included. In fact, the shape of this curve closely resembles the saturation curve of the MOSFET, providing strong evidence that the MOSFET is responsible for this nonlinear phenomenon.

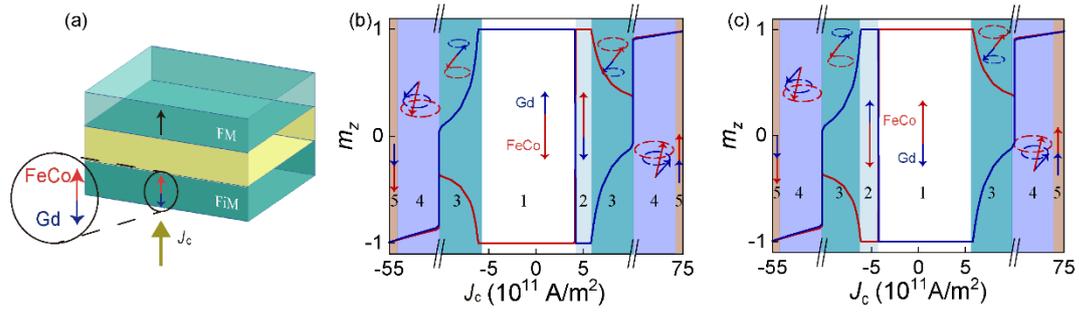

**Figure 1**. (a) Structure of the FiM-MTJ. (b) The phase diagram of the sample Gd$_{17.8}$FeCo$_{82.2}$ under STT starting in AP state. The switching region locates between $J_c$ = 4.2×10$^{11}$ and 5.8×10$^{11}$ A/m². (c) The phase diagram of the sample Gd$_{17.8}$FeCo$_{82.2}$ under STT starting in P state. The switching region locates between $J_c$ = -4.2×10$^{11}$ and -6.1×10$^{11}$ A/m².

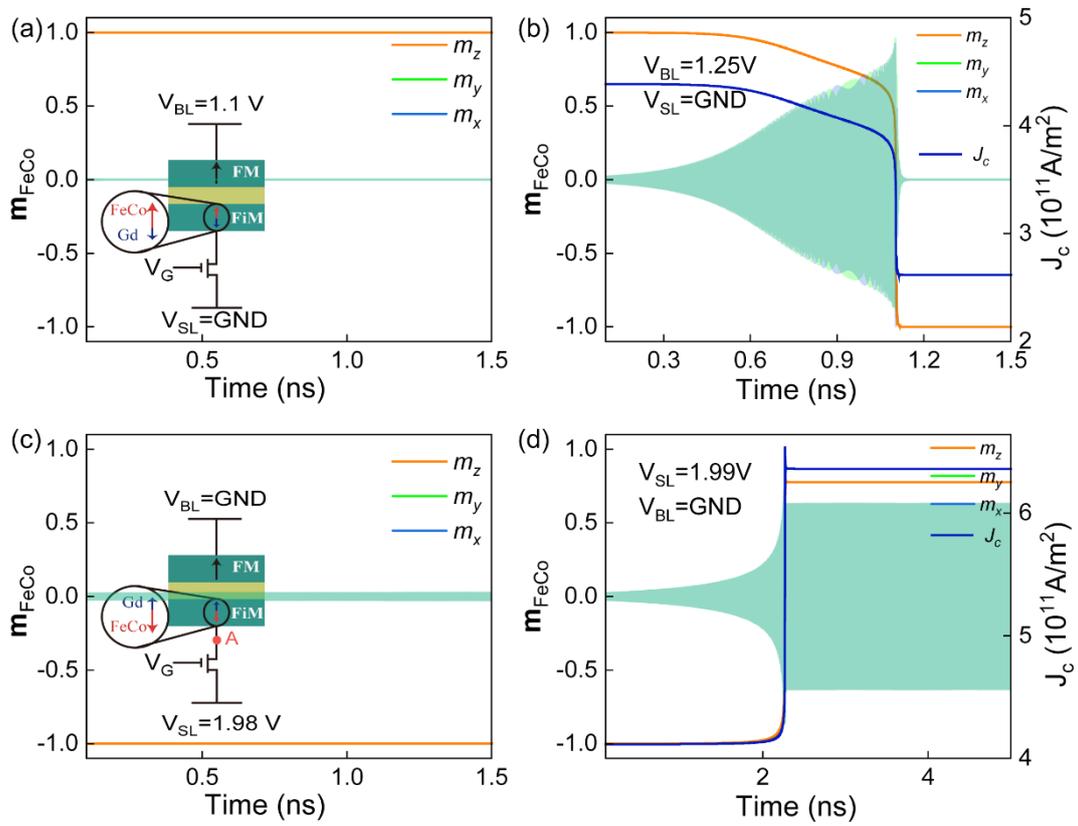

**Figure 2**. Starting from P state, the time evolution of **m**$_{FeCo}$ under (a) $V_{BL}$ = 1.1 V and (b) $V_{BL}$=1.25 V. Starting from AP state, the time evolution of **m**$_{FeCo}$ under (c) $V_{SL}$ = 1.98 V and (d) $V_{SL}$=1.99 V. The TMR ratio here is 70%.

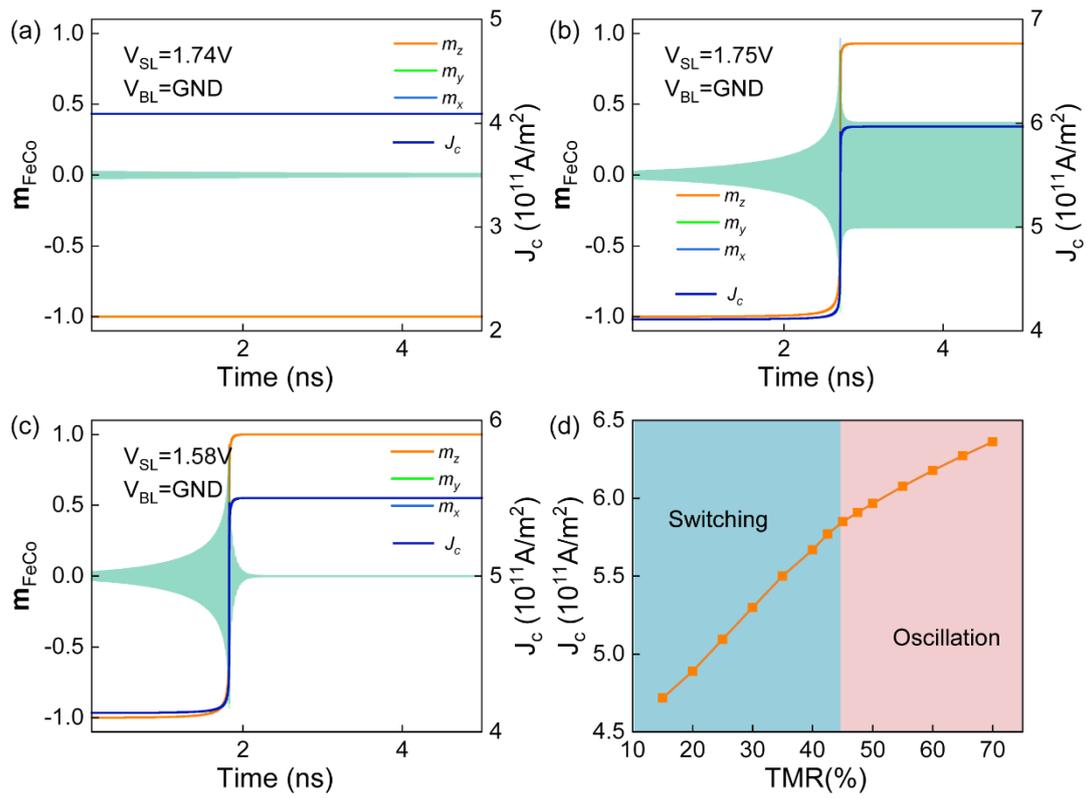

**Figure 3**. In the FiM-MTJ with TMR ratio of 50%, the time evolution of **m**$_{FeCo}$ under (a) $V_{SL}$ = 1.74 V and (b) $V_{SL}$=1.75 V. (c) In the FiM-MTJ with TMR ratio of 35%, the time evolution of **m**$_{FeCo}$ under $V_{SL}$ = 1.58 V. (d) The critical current density, over which the switching or oscillation occurs, as a function of TMR ratio.

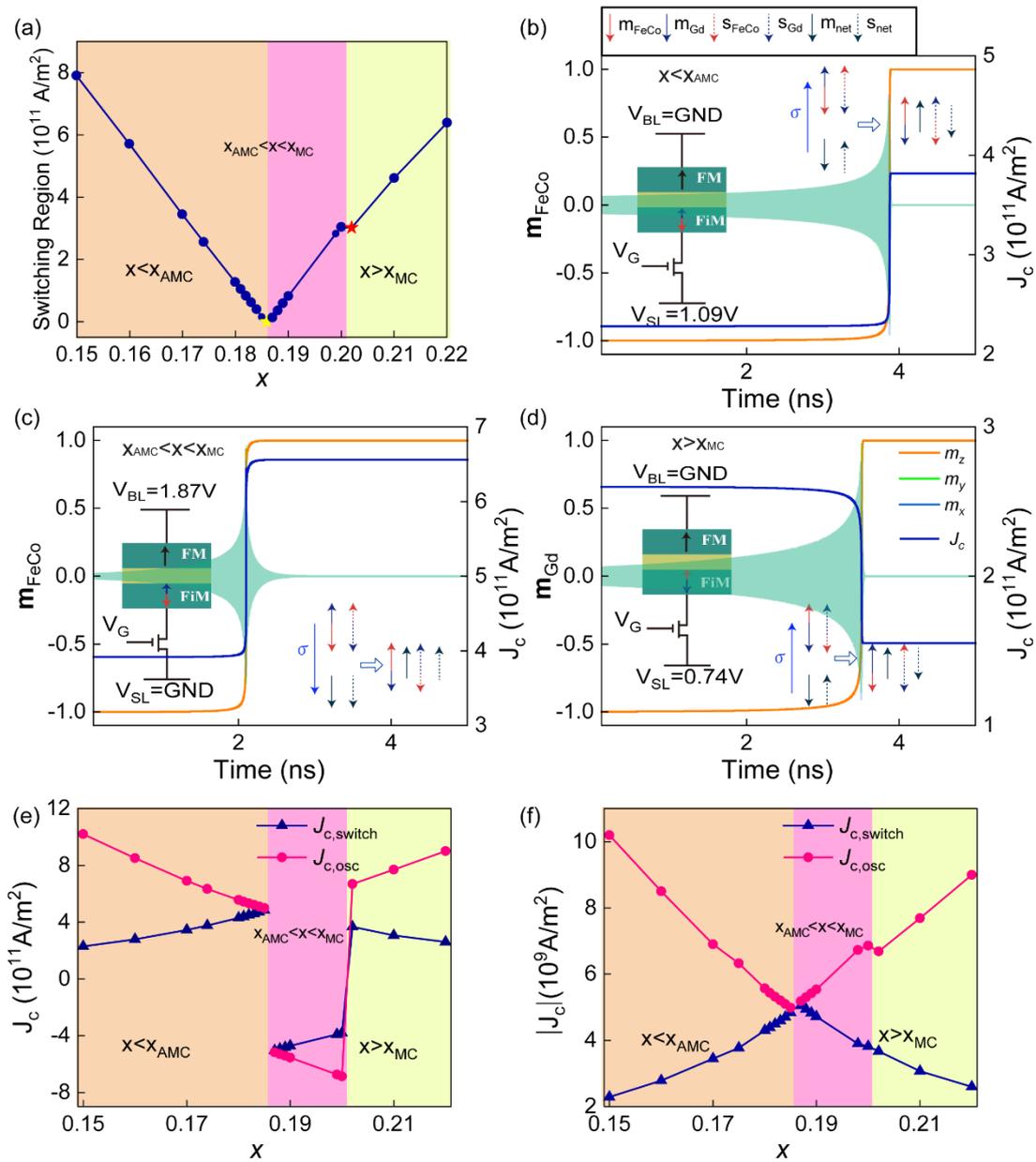

**Figure 4**. (a) Switching region as a function of $x$. Switching dynamics for (b) $x$ = 0.15, (c) $x$ = 0.199, and (d) $x$ = 0.22. These three samples are in $x<x_{AMC}$, $x_{AMC}<x<x_{MC}$, and $x>x_{MC}$, respectively. (e) $J_{c,switch}$ and $J_{c,osc}$ as a function of $x$. (f) The magnitude of $J_{c,switch}$ and $J_{c,osc}$ as a function of $x$.

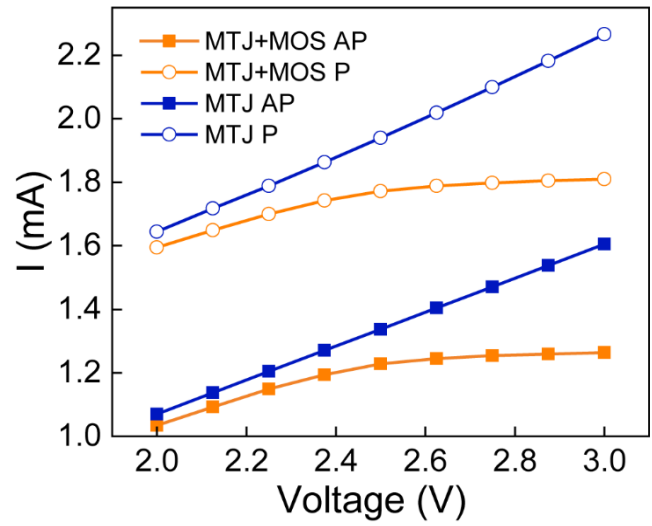

**Figure 5.** Current-voltage curves in the structure with and without MOSFET. The initial AP state is marked with filled square symbols, and the P state is indicated by empty circle symbols.